%% file: EXO-22-022_temp.tex
\pdfoutput=1
\documentclass[11pt,twoside,a4paper,cmspaper,final,collab]{cms-tdr}
\def\svnVersion{ea7658a}\def\svnDate{2025/02/03}\def\cmsCernNoTag{CERN-EP-2024-101}\def\cmsCernDate{\today}\def\cmsMessage{Published in Physical Review Letters as \href{http://dx.doi.org/10.1103/PhysRevLett.134.041801}{\doi{10.1103/PhysRevLett.134.041801}.}}
\begin{document}\cmsNoteHeader{EXO-22-022}

\newcommand{\masym}{\ensuremath{m_{\text{asym}}}\xspace}
\newcommand{\areco}{\ensuremath{\alpha^{\text{reco}}}\xspace}
\newcommand{\riso}{\ensuremath{r_{\text{iso}}}\xspace}
\newcommand{\mg}{\ensuremath{m_{\Gamma}}\xspace}
\newcommand{\mgone}{\ensuremath{m_{\Gamma_1}}\xspace}
\newcommand{\mgtwo}{\ensuremath{m_{\Gamma_2}}\xspace}
\newcommand{\mGG}{\ensuremath{m_{\Gamma\Gamma}}\xspace}
\newcommand{\mhat}{\ensuremath{\hat{m}_{\Gamma}}\xspace}
\newcommand{\pgg}{\ensuremath{\mathcal{D}_{\PGg\PGg}}\xspace}
\newcommand{\mPX}{\ensuremath{m_{\PX}}\xspace}
\newcommand{\mPGf}{\ensuremath{m_{\PGf}}\xspace}
\newcommand{\signal}{\ensuremath{\PX \to \PGf\PGf \to (\PGg\PGg)(\PGg\PGg)}\xspace}
\newcommand{\mxnf}{\ensuremath{(\mPX N)/f}\xspace}

\cmsNoteHeader{EXO-22-022}
\title{Search for new resonances decaying to pairs of merged diphotons in proton-proton collisions at \texorpdfstring{$\sqrt{s} = 13\TeV$}{sqrt(s) = 13 TeV}}

\date{\today}

\abstract{
A search is presented for an extended Higgs sector with two new particles, \PX and \PGf, in the process \signal. Novel neural networks classify events with diphotons that are merged and determine the diphoton masses.  The search uses LHC proton-proton collision data at $\sqrt{s} = 13\TeV$ collected with the CMS detector, corresponding to an integrated luminosity of 138\fbinv. No evidence of such resonances is seen.  Upper limits are set on the production cross section for \mPX between 300 and 3000\GeV and $\mPGf/\mPX$ between 0.5 and 2.5\%, representing the most sensitive search in this channel.
}

\hypersetup{
pdfauthor={CMS Collaboration},
pdftitle={Search for new resonances decaying to pairs of merged diphotons in proton-proton collisions at sqrt(s) = 13 TeV},
pdfsubject={CMS},
pdfkeywords={CMS, BSM, photons}}

\maketitle 

Many theories of particle physics beyond the standard model (BSM) predict the existence of extended Higgs sectors~\cite{th1,th4,th2,th5,th3,th6,th7}. Such extensions may include an approximate global symmetry with spontaneous symmetry breaking, which is a sufficient condition for the existence of new spin-0 particles, \PX and \PGf, with unknown masses \mPX and \mPGf. As long as $\mPX > 2\mPGf$, $\PX \to \PGf\PGf$ is an allowed decay mode. The \PGf boson can couple to pairs of standard model (SM) particles. Thus, the \PGf decay mode can result in various final state signatures, each presenting different experimental challenges, depending on the targeted kinematic regime. The CMS Collaboration previously performed a search targeting $\PX \to \PGf\PGf$, considering $\PGf \to \PQb\PAQb$ decays for \PGf boson masses above 25\GeV~\cite{Xaa4b}. Analogously to diphoton decays of the Higgs boson, the \PGf boson can decay to pairs of photons, $\PGf \to \PGg\PGg$, via loops of SM or BSM particles~\cite{originalMerged}.  The $\PGf \to \PGg\PGg$ channel provides sensitivity in the region of \PX and \PGf mass combinations that were beyond the reach of the $\PX \to \PGf\PGf \to (\PQb\PAQb)(\PQb\PAQb)$ search described in Ref.~\cite{Xaa4b}, including the region where the $(\PQb\PAQb)(\PQb\PAQb)$ channel is kinematically allowed but not well reconstructed. Prior searches for similar new spin-0 particles decaying to two photons were performed by the ATLAS Collaboration~\cite{ATLAS3g, ATLASgjets, ATLAS:2023ian}.  Searches for exotic decays of the Higgs boson have been performed by the CMS Collaboration; Ref.~\cite{CMS:2022xxa} searched for a four-photon final state and Ref.~\cite{CMSh4g} investigated the regime in which extremely collimated photon pairs are reconstructed as a single photon by the standard CMS algorithm.

This letter presents a search for the production of an \PX boson, decaying to a pair of \PGf bosons that themselves decay to pairs of photons, \signal. The analysis focuses on a broad range of \mPX between 0.3 and 3\TeV, with a relative mass $\alpha=\mPGf/\mPX$ ranging between 0.5 and 2.5\% corresponding to events for which the angular separation ($\Delta R$) of the photons in a pair is between 0.01 and 0.2 radians. In this range, the separation is too small for the photons to be individually identified in the CMS detector, but not so small that the signature would be consistent with a single photon in the standard CMS reconstruction~\cite{CMS_PHO_ID}. These mass combinations correspond to the regime where the previous $\PGf \to \PQb\PAQb$ search has no sensitivity. To gain sensitivity in this intermediate regime, a dedicated convolutional neural network (CNN) has been developed, which classifies energy clusters in the CMS electromagnetic calorimeter (ECAL) as originating from two overlapping photons (denoted $\Gamma$), hadronic activity, or a single photon. A second CNN is used to reconstruct the mass of the $\Gamma$ candidates, similar to~\cite{CMSetoeML}, optimized for the exploration of the $\alpha$--\mPX space considered in this letter.  After selecting events with two $\Gamma$ candidates, the analysis searches for localized excesses over the steeply falling background distribution of their paired reconstructed mass (\mGG).  The search is performed by simultaneously fitting the \mGG spectrum in bins of the reconstructed mass ratio $\areco=\mhat/\mGG$, where \mhat is the average reconstructed mass of the two $\Gamma$ candidates.  The \areco binning is chosen based on the resolution obtained from the simulated signal samples. The signals may appear in multiple \areco bins.  The background in this search is derived exclusively from a parameterized fit to the data.

The search uses a data set of proton-proton ($\Pp\Pp$) collisions at $\sqrt{s}=13\TeV$ at the CERN LHC, collected with the CMS detector in 2016--2018, corresponding to an integrated luminosity of 138\fbinv.  The CMS apparatus~\cite{CMS:2008xjf} is a multipurpose, nearly hermetic detector, designed to trigger on~\cite{CMS:2020cmk,CMS:2016ngn} and identify electrons, muons, photons, and hadrons~\cite{CMS:2020uim,CMS:2018rym,CMS:2014pgm}. A global reconstruction algorithm~\cite{CMS:2017yfk} combines the information provided by the all-silicon inner tracker and by the lead tungstate ECAL and brass-scintillator hadron calorimeters, operating inside a 3.8\unit{T} superconducting solenoid, with data from gas-ionization muon detectors embedded in the solenoid return yoke, to build \PGt leptons, jets, missing transverse momentum, and other physics objects~\cite{CMS:2018jrd,CMS:2016lmd,CMS:2019ctu}. Events are retained for analysis if they pass a trigger requiring two photon candidates, each with $\pt > 60$ or 70\GeV in 2016 or 2017--2018 data, respectively. The photon identification algorithm used by the trigger only selects photons based on isolation, and not on the cluster shape, and hence has high efficiency for selecting events with two merged diphotons, provided they have sufficient \pt.  Tabulated results for this analysis are provided in the HEPData record~\cite{hepdata}.

The ECAL barrel covers a pseudorapidity range $\abs{\eta} < 1.4$ and consists of 61,200 individual crystals, each covering $\Delta\eta{\times}\Delta\phi = 0.0175{\times}0.0175$ where $\phi$ is the azimuthal angle.  Individual photons typically deposit energy in several adjacent crystals, and are reconstructed as localized clusters of energy. For merged diphotons, such clusters may overlap.  Purpose-built algorithms for identifying diphotons from the shapes of their energy distributions in the ECAL barrel are described below. To simplify the CNN implementation, this analysis uses events where both $\Gamma$ candidates are reconstructed in the barrel, which account for more than 60\% of the expected signal yield.

The benchmark signal model \signal, produced via gluon fusion in the narrow-width approximation, is generated at leading order with the \MGvATNLO 2.6.0~\cite{madgraph} generator, with up to two additional partons in the matrix element calculations. Simulated signals are generated for \mPX from 0.3 to 3\TeV (at intervals ranging from 0.1 to 0.5\TeV) and for $\alpha$ from 0.5 to 2.5\% (at 0.5\% intervals). The signal production cross section is calculated numerically at next-to-next-to-leading order (NNLO) using the \textsc{HqT} 2.0 program~\cite{theory1,theory2,theory3}, and depends on the number ($N$) of new Dirac quarks that receive all their mass from the \PX vacuum expectation value $f$, so that the signal cross section depends on the value \mxnf. The background from SM production of jets and photons ($\PGg\text{+jets}$), used to optimize the analysis procedure, is also modeled at next-to-leading order with the \MGvATNLO generator. Hadronic objects for training of the classification CNN are taken from this sample. The parton showering and fragmentation for these samples is modeled by \PYTHIA 8.240~\cite{pythia8}. Matching between the matrix element and parton shower jets relies on the MLM matching procedure~\cite{mlmmatch}. Samples of isolated photons and diphotons with uniform energy distributions are generated with \PYTHIA 8.107~\cite{pythia81} and are used to train the $\Gamma$ identification and mass reconstruction CNNs.  The CP5~\cite{pythiaTune,m_2020_extraction} underlying event tune is used for the simulation with the NNPDF3.1~\cite{nnpdf} NNLO parton distribution function sets. The response of the CMS detector is modeled using \GEANTfour~\cite{geant}.  The effect of extra $\Pp\Pp$ interactions in the same or adjacent bunch crossings (pileup) is incorporated in the simulation, with the frequency distribution of additional vertices adjusted to match that observed in data.

Energy deposits in the ECAL barrel are clustered into $\Gamma$ candidates using an algorithm inspired by the Cambridge--Aachen algorithm~\cite{CAjets}.  Beginning with the most energetic single-crystal deposit in the ECAL barrel, the nearest neighbor energy deposit is iteratively added to the cluster if the $\Delta R$ between the deposit and the energy-weighted average position of the current cluster is less than a radius $R=0.15$, where $\Delta R = \sqrt{\smash[b]{\Delta\eta\,^2 + \Delta\phi\,^2}}$.  When no new crystals can be included in a cluster, the process begins again with the highest energy unclustered deposit. The distance parameter was chosen to be large enough to capture both photons from a single \PGf boson decay in a single cluster, but not so large that the usual CMS photon identification would reconstruct two separate photons. For the range of $\alpha$ values considered in this analysis, the simulated efficiency of this algorithm to capture both photons in the cluster ranges from 80--100\%, while the efficiency of requiring two individually identified photons ranges from 1 to 20\%.

Each $\Gamma$ candidate consists of a collection of energy deposits on an array of crystals, and can be represented as a pixelated image. Convolutional neural networks developed for image recognition are well suited to extract relevant information about the $\Gamma$ candidates.  The clustered energy deposits are converted to a $30{\times}30$ image, centered on the lower left corner of the crystal with the most energetic deposit, where the brightness of a pixel corresponds to the energy deposited in a single crystal.  Two different CNNs are used. The first separates $\Gamma$ candidates arising from  merged diphotons from those produced by hadronic decays and single photons. The second regresses the mass-to-energy ratio ($m/E$) of the $\Gamma$ candidate, under the assumption that it was produced by a genuine diphoton. The architectures of both CNNs are based on the \textsc{DeepTop}~\cite{Macaluso:2018tck} CNN, modified as described below.

For both CNNs, the inputs are the individual crystal energies of that $\Gamma$ candidate, normalized such that the sum of all crystal energies is unity. These are passed through a series of four convolutional layers with kernels of sizes $7{\times}7$, $5{\times}5$, $4{\times}4$, and $3{\times}3$, with 64 output feature maps after each layer in the classification network and 16 in the mass regression network.
For the classification, this is followed by one fully-connected (FC) linear layer with three outputs corresponding to the output classes. A logarithmic softmax function is used to convert these to classifier scores, ranging from 0 to 1, with higher scores indicating that the cluster is more likely to belong to each category: diphoton (\pgg), photon ($\mathcal{D}_{\PGg}$), or hadron ($\mathcal{D}_\text{had}$), with $\pgg + \mathcal{D}_{\PGg} + \mathcal{D}_\text{had} = 1$.  For the mass regression, three FC layers are used with output sizes of 64, 16, and 1, respectively, and the geometric $\eta$ of the $\Gamma$ candidate is used as an additional input to the first FC layer. The output of the regression CNN is a single number corresponding to the predicted value of $m/E$.

The training of the classification CNN uses 600k events, divided equally between the diphoton, single photon, and hadron categories. For the regression, 500k simulated diphoton events are used. In both cases, diphoton events are sampled to retain a flat $m/E$ spectrum from 0 to 0.07. The loss function used for classification is the categorical cross-entropy function; for the regression, a modified mean squared error function optimized for small positive values is used: $\sum_{i} -\log{( \re^{-(p_i - q_i)^2} )} + \log{( 1-\re^{-(p_i - q_i)^2} )}$, where $p_i,q_i$ are the true and predicted $m/E$ values for the $i$th event, respectively.

To suppress the dominant background from misidentified jets, a relative isolation \riso is defined and applied to each $\Gamma$ candidate in the analysis. Jets reconstructed with the anti-\kt algorithm~\cite{Cacciari:2008gp} and a distance parameter of 0.4, using the \textsc{FastJet} package~\cite{ref:fastjet}, are used. If a $\Gamma$ candidate overlaps with such a jet within $\Delta R < 0.15$, its \riso is defined as the ratio of its energy to that of the jet. Otherwise, it is set to unity.

\begin{figure*}[htb!]
\centering
  \includegraphics[width=0.49\textwidth]{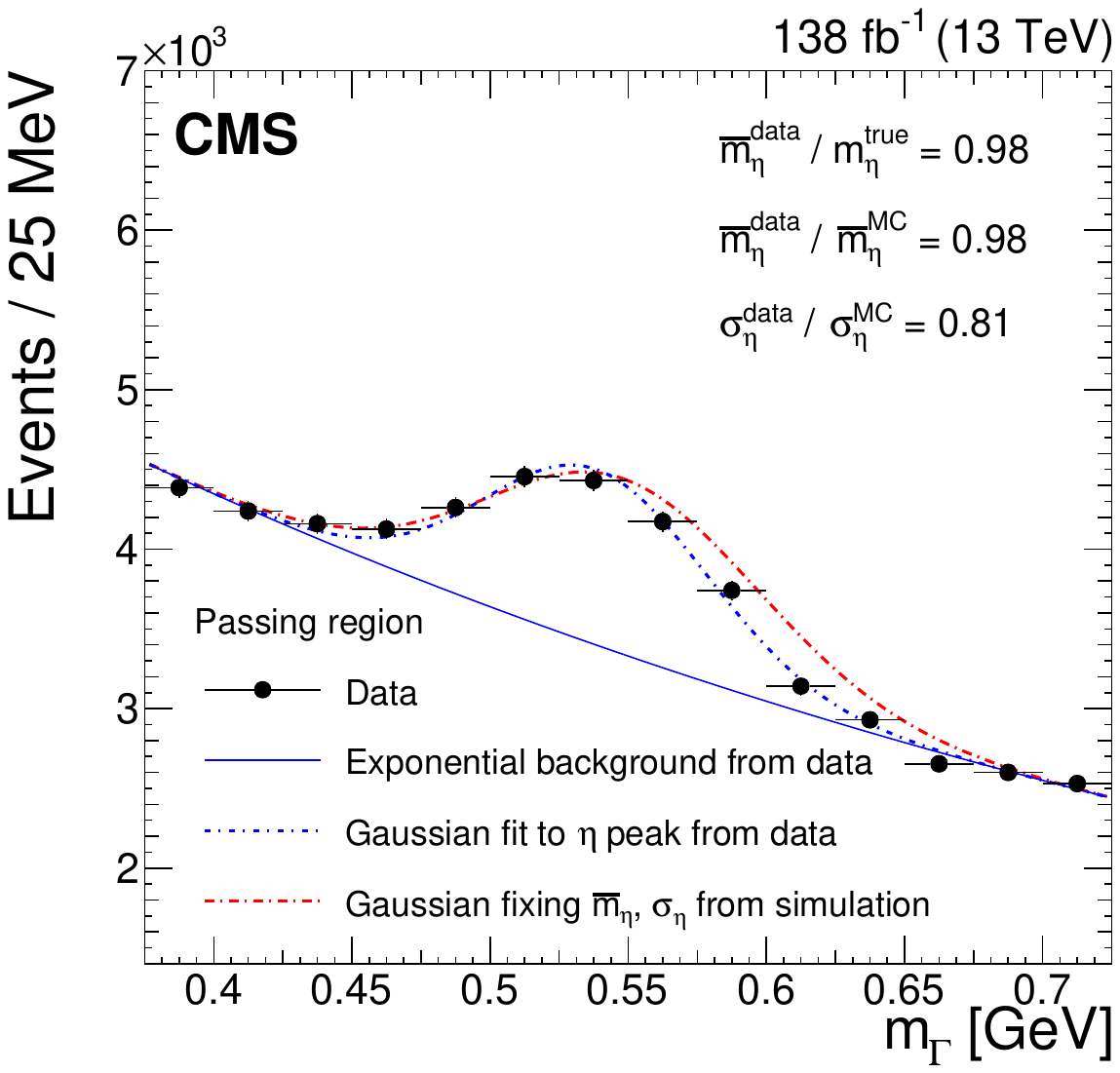}
  \includegraphics[width=0.49\textwidth]{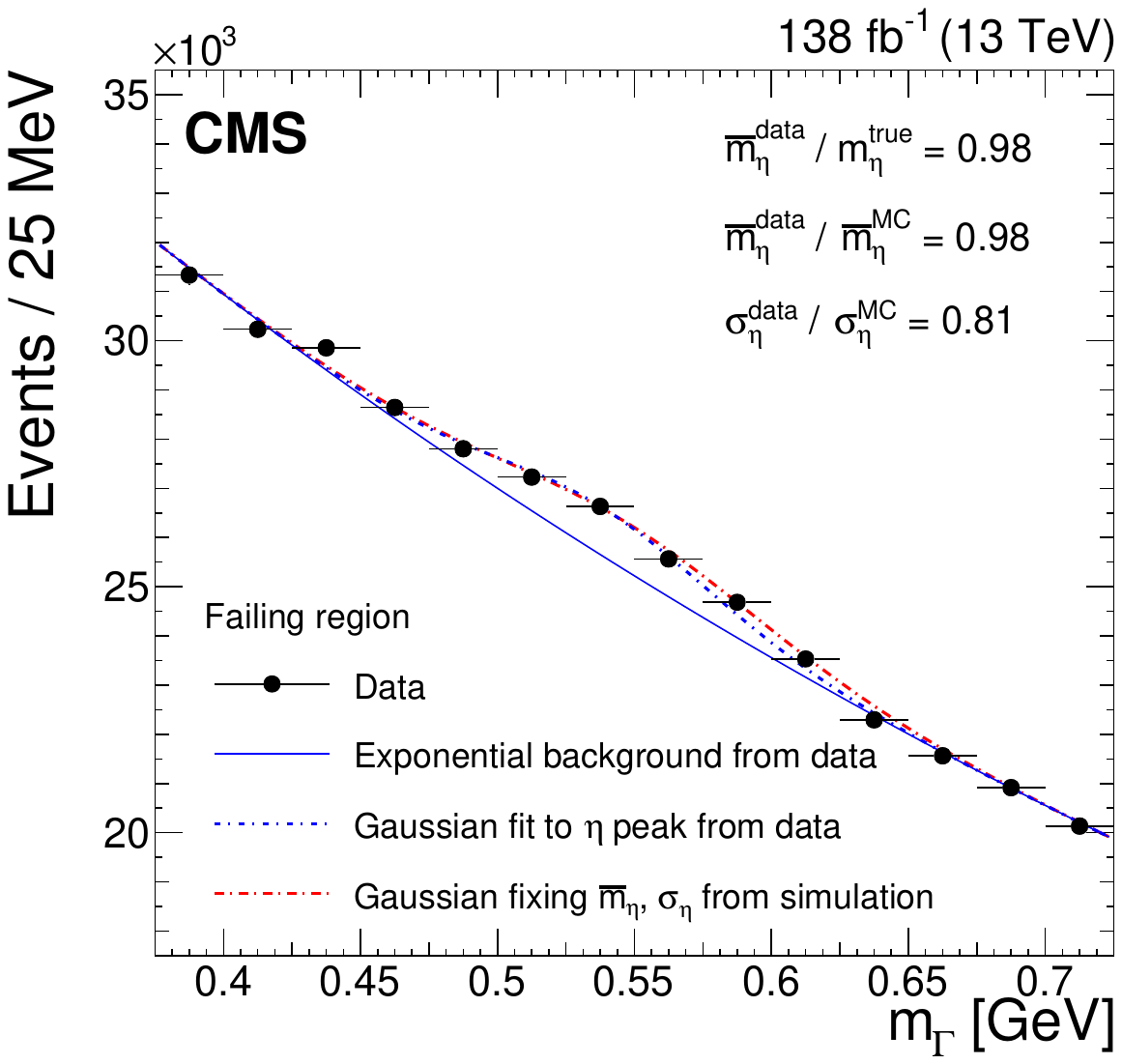}
  \caption{Cluster mass (\mg) distribution in data for both the passing (left) and failing (right) regions, in the energy range for which the \Pgh meson is expected to form a single $\Gamma$ candidate. The signal (background) is modeled by a Gaussian (exponential) function. Blue and red dashed lines depict Gaussian fits to the data and Monte Carlo (MC) simulation, respectively. The solid blue line shows the background component of the fit. Ratios of the Gaussian fit means ($\overline{m}$) and widths ($\sigma$) are displayed, where $m_{\Pgh}^{\text{true}}$ is the true mass of the \Pgh meson.}
\label{fig:etas}
\end{figure*}

The diphoton decays of \Pgpz or \Pgh mesons copiously produced in hadronic showers can be used to validate the performances of the CNNs. While the technique developed in Ref.~\cite{CMSetoeML} utilizes \Pgpz  decays to validate its mass regression, here the \Pgh meson is used as, in the merged signature, it has a similar Lorentz boost to the considered signals. An \Pgh meson control sample is selected to validate the efficiency of the signal classification and the scale and resolution of the $m/E$ regression.  The control sample is taken from all events passing an electron or photon trigger.  The $\Gamma$ candidates are required to have energies in the range 30--60\GeV and $\riso>0.5$. A study made with simulated data shows that the performance of the CNNs does not depend on the energy or mass of the diphotons. This energy range is selected such that the \Pgh meson decay products form a single $\Gamma$ candidate, providing a direct validation of a subset of the $m/E$ range for which the CNNs were trained. The \mg distributions are compared in two subsets, depending on whether $\pgg>0.9$ or not (referred to as ``passing'' and ``failing'' regions).  The \mg distributions of the passing and failing regions are presented in Fig.~\ref{fig:etas}, showing a distinct peak at the \Pgh meson mass, providing strong evidence that the regression CNN is working as intended. Due to contamination from other particles produced in the same jet, a substantial number of \Pgh mesons decaying to collimated diphotons will appear in the failing region, resulting in the peak seen there.  The data and the simulation are each fit simultaneously in the passing and failing regions; the mass of the \Pgh meson in both is reconstructed to within 2\% of its true value.  The classifier efficiency and the mean and width of the peak are extracted using a fit to data, where the signal (background) is modeled by a Gaussian (exponential) function.   The classifier efficiency in the \Pgh meson control region, extracted from the relative yields in the passing and failing distributions, is measured to be 55 (53)\% in data (simulation).  The data-to-simulation efficiency ratio is consistent with unity, with a total uncertainty of 10\% that is propagated to the final results. The systematic uncertainty in the \mg resolution is obtained from the fractional difference in the widths of the \Pgh meson peaks in data and simulation, corresponding to 23\%, which is correlated across all \areco bins.  Finally, the mean positions of the peaks are found to be consistent within the respective statistical uncertainties.

We further validate the performance of the regressor and classifier using a control sample of reconstructed \PZ bosons through the decay $\PZ \to \Pep \Pem$ by exploiting the case when each electron undergoes bremsstrahlung and appears consistent with a merged diphoton signature.  Events are selected by requiring two clusters with $\pgg>0.75$ and $\riso>0.8$.  The resulting reconstructed \mGG distribution shows a clear presence of \PZ bosons in both data and simulation, with nearly identical performance. The positions of the reconstructed mass peaks in data and simulation are found to be within 0.5\% across a wide range of cluster \pt, and this value is applied as a systematic uncertainty in the $\Gamma$ candidate energy scale.

The offline analysis considers events with two $\Gamma$ candidates in the ECAL barrel with $\pt>90\GeV$, where the trigger becomes fully efficient. Both $\Gamma$ candidates must also fulfill the following requirements: $\riso>0.8$ and $\pgg>0.9$. This \pgg requirement retains $\gtrsim$80\% of simulated signal across all masses and rejects approximately 99\% of the background. The mass asymmetry of the event $\masym = \abs{\mgone - \mgtwo}/(\mgone + \mgtwo)<0.25$ is required. Finally, $\Delta\eta<1.5$ (between the candidates) is required to further suppress background from the SM production of jets and photons.

\begin{figure}[ht]
\centering
  \includegraphics[width=0.49\textwidth]{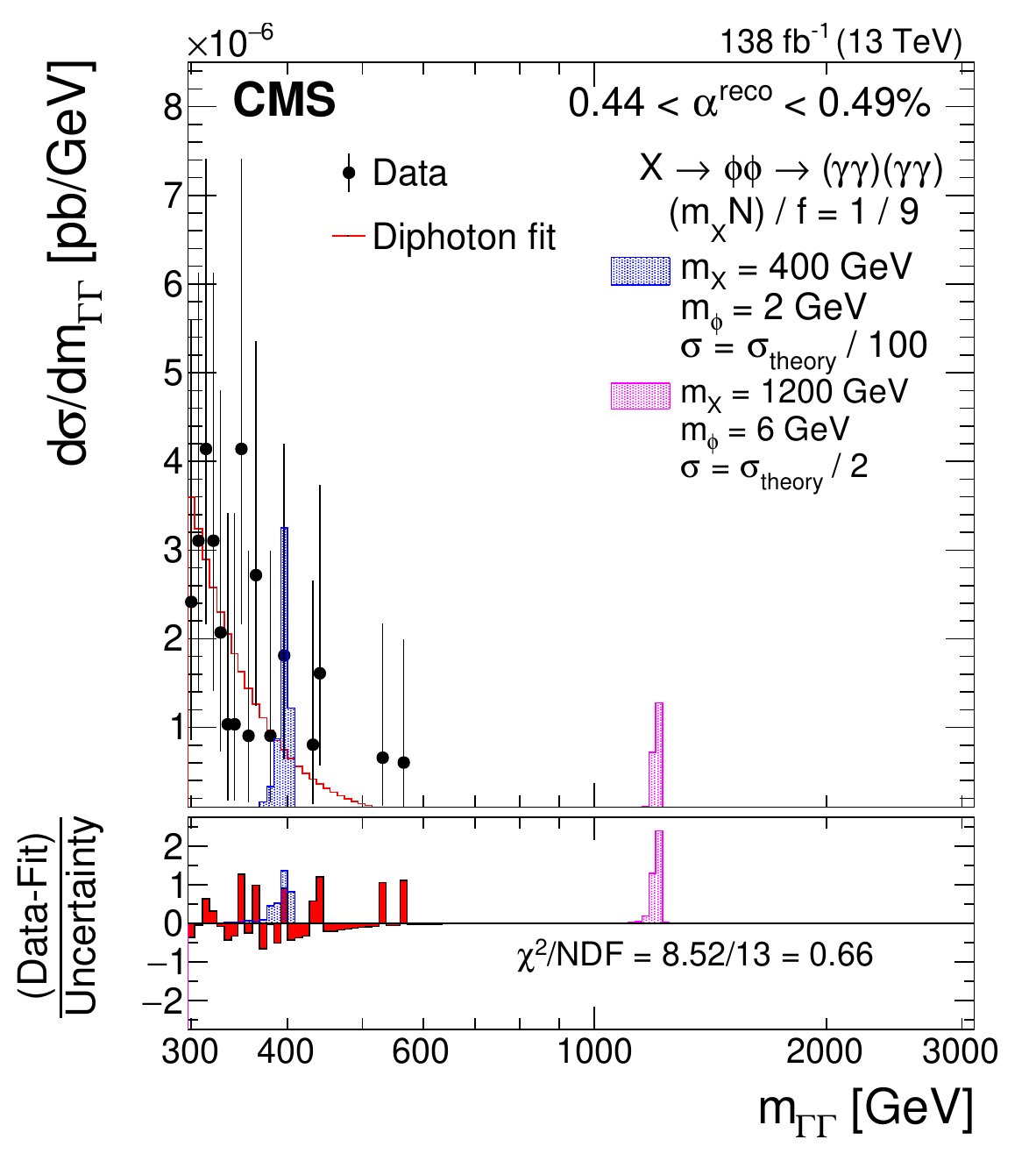}
  \caption{Dicluster mass (\mGG) distribution for the data (points) for one of the \areco bins of the search ($0.44<\areco<0.49\%$ ), fitted with the diphoton function (red), one of the considered five background parametrizations. Examples of two representative predicted signals are shown (blue and pink). The lower panel shows the difference between the observed data and the background prediction divided by the statistical uncertainty of the data ($\sigma_{\text{data}}$), the aforementioned signals divided by $\sigma_{\text{data}}$, and the goodness of fit measure $\chi^2/\text{NDF}$ (where NDF is the number of degrees of freedom).}
\label{fig:data_bkg_fit1}
\end{figure}

The remaining data are evaluated for localized excesses in the \mGG distribution, in nine non-overlapping divisions of the \areco distributions ranging from 0.3 to 3\%, each determined by a combination of detector resolution and a requirement that the division contains enough events for the background estimate to converge. Any particular signal is expected to appear in only a few adjacent \areco divisions. As in previous searches~\cite{Sirunyan:2018xlo,Sirunyan:2016iap,Khachatryan:2015dcf,ATLAS:2015nsi}, the background shape is modeled by fitting empirical functional forms to the observed data in each division, where the number of parameters in the functional forms is determined by performing Fisher F-tests~\cite{Ftest} on progressively higher-order functions. 
Functions with three parameters were found to be acceptable for describing the data: the dijet function $p_0  (1-x)^{p_1} / x^{p_2}$, the modified dijet function $p_0 (1-x^{1/3})^{p_1} / x^{p_2}$, the diphoton function $p_0  x^{p_1 + p_2 \log{x}}$, a power-law times an exponential function $p_0 \re^{-p_1 x} / x^{p_2} $, and a four-parameter power-law function $p_0  p_1^{p_2 x + p_3/x}$, where $x$ is $\mGG/\sqrt{s}$.  The choice of function is encoded into the fitting procedure as a discrete parameter of the likelihood. Signal shapes are modeled by fitting a double-sided Crystal Ball function~\cite{CrystalBallRef} to the reconstructed simulated signal \mGG spectra, then interpolating the function parameters to generate shapes with fine spacing in \mPX.  To validate the robustness of the fit, a goodness-of-fit test and bias tests are performed. The bias tests use simulated events with a variety of simulated signals injected. No significant bias is observed for any \PX and \PGf boson mass combination. The result of the fit for one representative \areco division is shown in Fig.~\ref{fig:data_bkg_fit1}. The search uses a fit of the background function plus the simulated signal shape to the data, taking into account statistical and systematic uncertainties, and is performed for $\mGG>297\GeV$ and simultaneously in all bins of \areco. 

Systematic uncertainties are modeled in the fit as nuisance parameters that affect the shapes and normalizations of signal and background processes, with log-normal priors for the uncertainties affecting only the normalization and Gaussian priors for those affecting the shapes of distributions. The main sources of systematic uncertainty in the background modeling are the choice of the background function and the background function fit parameters.  The parameters of the background function are treated as freely floating nuisance parameters, and are evaluated via profiling.  The discrete profiling method~\cite{Dauncey:2014xga} is used for considering the choice of the functional form as a discrete nuisance parameter,  which is profiled in an analogous way to continuous nuisance parameters.  Systematic uncertainties affecting the predicted signal yield include the uncertainty in the integrated luminosity measurement (1.6\%)~\cite{CMS-LUM-17-003,CMS-PAS-LUM-17-004,CMS-PAS-LUM-18-002},  the diphoton classifier efficiency (10\%, applied per cluster), the \mg resolution uncertainty (23\%, affecting the yield in each \areco bin), and the trigger efficiency (5\%). Systematic uncertainties affecting the shape of the predicted signal distributions include the $\Gamma$ energy scale uncertainty (0.5\%, applied per cluster) and uncertainties in the modeling of pileup (1--10\%). The dominant systematic uncertainty in the signal modeling is due to the diphoton classifier efficiency.

\begin{figure}[th]
\centering
  \includegraphics[width=.49\textwidth]{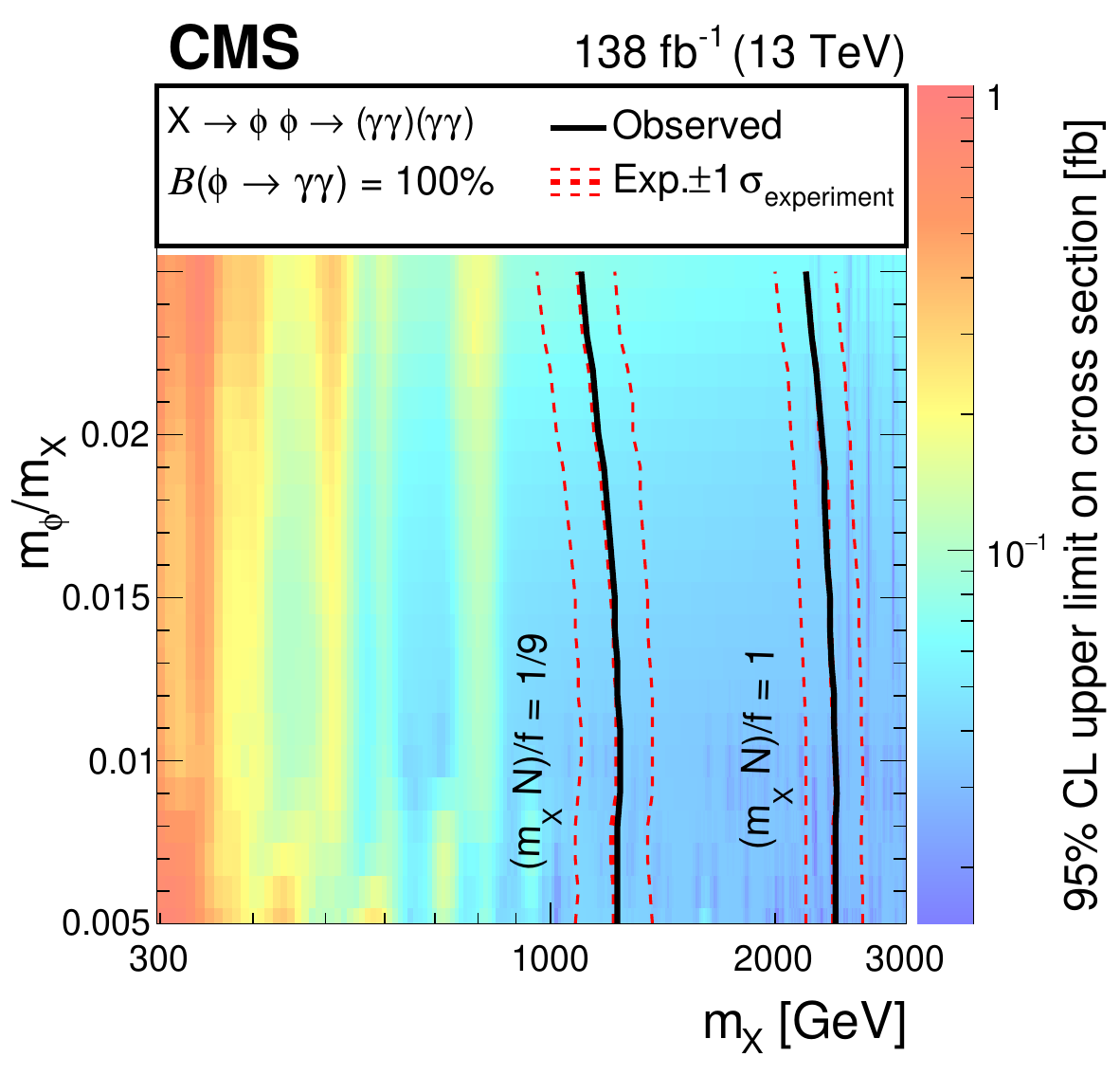}
  \caption{Exclusion limits at 95\% \CL on $\sigma(\signal)$ displayed in the ($\mPGf/\mPX$)--\mPX plane. Branching fractions ($\mathcal{B}$) of both $\PX \to \PGf\PGf$ and $\PGf \to \PGg\PGg$ are assumed to be 100\%. The black (red) lines represent the observed (expected) mass exclusions corresponding to different assumptions of \mxnf. The observed upper limits on the cross section are shown on the color $z$ axis.}
\label{fig:limitsTwod}
\end{figure}

No significant excesses are observed.  The largest excess corresponds to $\mPX \approx 720\GeV$ and $\alpha =0.7\%$ ($\mPGf \approx 5\GeV$) with a local (global) significance of 3.57 (1.07) standard deviations.  The global significance accounts for the look-elsewhere effect~\cite{abc} by using pseudo-experiments to compute the probability that the background hypothesis produces a signal-like fluctuation with at least the observed local significance anywhere in the sensitive range of \mPX and $\alpha$. The fit results are used to set 95\% confidence level (\CL) upper limits on the cross section $\sigma(\Pp \Pp \to \PX)$, assuming a 100\% branching fraction for \signal.  Upper limits are computed using a modified frequentist approach, based on the \CLs criterion~\cite{CLS2,CLS1,CMS:2024Combine}, with the profile likelihood ratio used as the test statistic within the asymptotic approximation~\cite{Cowan:2010js}.  Observed and expected limits are computed as functions of \mPX for a given assumed value of $\alpha$, and compared to the theoretical estimates of $\sigma(\PX \to\PGf\PGf)$ for a set of $(\mPX N)/f$ values.  Figure~\ref{fig:limitsTwod} shows the observed upper limits on the cross section and the mass exclusion curves in the $\alpha$--\mPX plane.  The observed (expected) upper limits on this process range from 0.03--1.06 (0.03--0.79)\unit{fb}, depending on \mPX and \mPGf.

In summary, a search for an extended Higgs sector with two new particles, \PX and \PGf, with unknown masses \mPX and \mPGf, has been presented for the decay sequence \signal.  The search uses proton-proton collision data at $\sqrt{s}=13\TeV$, collected with the CMS detector at the LHC in 2016--2018, corresponding to an integrated luminosity of 138\fbinv.  The analysis considers \mPX between 0.3 and 3\TeV, and is restricted to values of \mPGf for which the ratio $\mPGf / \mPX$ is between 0.5 and 2.5\%.  As a result, the two photons from each \PGf boson overlap significantly in the electromagnetic calorimeter.  Convolutional neural networks trained on clusters of calorimeter energy deposits are used to classify events containing merged diphotons and to regress the mass of the diphoton system.  The dicluster mass spectra, in bins of the ratio of the average cluster mass divided by the dicluster mass, are analyzed for the presence of new resonances, and are found to be consistent with the standard model expectations. Upper limits are set at 95\% confidence level (\CL) on the production cross section for \signal, as a function of the resonance masses, where both the $\PX \to \PGf\PGf$ and $\PGf \to \PGg\PGg$ branching fractions are assumed to be 100\%. Observed (expected) limits range within 0.03--1.06 (0.03--0.79)\unit{fb} at 95\% \CL for the masses considered. For the masses considered, these results represent the most sensitive search for extended Higgs sectors that have two new resonances decaying to the four-photon final state.

\begin{acknowledgments}
  We congratulate our colleagues in the CERN accelerator departments for the excellent performance of the LHC and thank the technical and administrative staffs at CERN and at other CMS institutes for their contributions to the success of the CMS effort. In addition, we gratefully acknowledge the computing centers and personnel of the Worldwide LHC Computing Grid and other centers for delivering so effectively the computing infrastructure essential to our analyses. Finally, we acknowledge the enduring support for the construction and operation of the LHC, the CMS detector, and the supporting computing infrastructure provided by the following funding agencies: SC (Armenia), BMBWF and FWF (Austria); FNRS and FWO (Belgium); CNPq, CAPES, FAPERJ, FAPERGS, and FAPESP (Brazil); MES and BNSF (Bulgaria); CERN; CAS, MoST, and NSFC (China); MINCIENCIAS (Colombia); MSES and CSF (Croatia); RIF (Cyprus); SENESCYT (Ecuador); ERC PRG, RVTT3 and MoER TK202 (Estonia); Academy of Finland, MEC, and HIP (Finland); CEA and CNRS/IN2P3 (France); SRNSF (Georgia); BMBF, DFG, and HGF (Germany); GSRI (Greece); NKFIH (Hungary); DAE and DST (India); IPM (Iran); SFI (Ireland); INFN (Italy); MSIP and NRF (Republic of Korea); MES (Latvia); LMTLT (Lithuania); MOE and UM (Malaysia); BUAP, CINVESTAV, CONACYT, LNS, SEP, and UASLP-FAI (Mexico); MOS (Montenegro); MBIE (New Zealand); PAEC (Pakistan); MES and NSC (Poland); FCT (Portugal); MESTD (Serbia); MCIN/AEI and PCTI (Spain); MOSTR (Sri Lanka); Swiss Funding Agencies (Switzerland); MST (Taipei); MHESI and NSTDA (Thailand); TUBITAK and TENMAK (Turkey); NASU (Ukraine); STFC (United Kingdom); DOE and NSF (USA).
\end{acknowledgments}

\bibliography{auto_generated}

\cleardoublepage \appendix\section{The CMS Collaboration \label{app:collab}}\begin{sloppypar}\hyphenpenalty=5000\widowpenalty=500\clubpenalty=5000\input{EXO-22-022-public-authorlist.tex}\end{sloppypar}
%%% END EDITABLE REGION %%%
% skeleton_end
\end{document}

%% file: EXO-22-022-public-authorlist.tex
\cmsinstitute{Yerevan Physics Institute, Yerevan, Armenia}
{\tolerance=6000
A.~Hayrapetyan, A.~Tumasyan\cmsAuthorMark{1}\cmsorcid{0009-0000-0684-6742}
\par}
\cmsinstitute{Institut f\"{u}r Hochenergiephysik, Vienna, Austria}
{\tolerance=6000
W.~Adam\cmsorcid{0000-0001-9099-4341}, J.W.~Andrejkovic, T.~Bergauer\cmsorcid{0000-0002-5786-0293}, S.~Chatterjee\cmsorcid{0000-0003-2660-0349}, K.~Damanakis\cmsorcid{0000-0001-5389-2872}, M.~Dragicevic\cmsorcid{0000-0003-1967-6783}, P.S.~Hussain\cmsorcid{0000-0002-4825-5278}, M.~Jeitler\cmsAuthorMark{2}\cmsorcid{0000-0002-5141-9560}, N.~Krammer\cmsorcid{0000-0002-0548-0985}, A.~Li\cmsorcid{0000-0002-4547-116X}, D.~Liko\cmsorcid{0000-0002-3380-473X}, I.~Mikulec\cmsorcid{0000-0003-0385-2746}, J.~Schieck\cmsAuthorMark{2}\cmsorcid{0000-0002-1058-8093}, R.~Sch\"{o}fbeck\cmsorcid{0000-0002-2332-8784}, D.~Schwarz\cmsorcid{0000-0002-3821-7331}, M.~Sonawane\cmsorcid{0000-0003-0510-7010}, S.~Templ\cmsorcid{0000-0003-3137-5692}, W.~Waltenberger\cmsorcid{0000-0002-6215-7228}, C.-E.~Wulz\cmsAuthorMark{2}\cmsorcid{0000-0001-9226-5812}
% ...
\par}
\vskip\cmsinstskip
\dag:~Deceased\\
$^{1}$Also at Yerevan State University, Yerevan, Armenia\\
$^{2}$Also at TU Wien, Vienna, Austria\\
% ...
